\documentclass[useAMS,useusegraphicx]{mn2e}
\usepackage{amssymb}
\usepackage{epsfig}

\newcommand{\mnras}{MNRAS}
\newcommand{\apj}{ApJ}
\newcommand{\aaps}{A\&AS}
\newcommand{\aap}{A\&A}
\newcommand{\aj}{AJ}
\newcommand{\apjl}{ApJ}
\newcommand{\apjs}{ApJS}
\newcommand{\nat}{Nature}
\newcommand{\araa}{ARA\&A}

\def\etal{et al.\ }

\def\gne #1#2{\ \vphantom{S}^{\raise-0.5pt\hbox{$\scriptstyle#1$}}_
{\raise0.5pt \hbox{$\scriptstyle#2$}}}

\def\araa#1{{\it Ann. Rev. Astron. Astrophys. }{\bf #1}}

\title[Luminosity dependent star-formation history of S0 galaxies]{Luminosity dependent star-formation history of S0 galaxies: evidence from GALEX-SDSS-2MASS-WISE colours}
\author[Barway \etal]
{Sudhanshu Barway,$^{1}$\thanks{E-mail: barway@saao.ac.za (SB)}
  Yogesh Wadadekar,$^{2}$\thanks{E-mail: yogesh@ncra.tifr.res.in (YW)}
  Kaustubh Vaghmare$^{3}$\thanks{E-mail: kaustubh@iucaa.ernet.in
   (KV)}
\newauthor
and Ajit K. Kembhavi$^{3}$\thanks{E-mail: akk@iucaa.ernet.in (AKK)}\\
$^{1}$South African Astronomical Observatory, P.O. Box 9, 7935, Observatory, Cape Town, South Africa\\
$^{2}$National Centre for Radio Astrophysics, Tata Institute of Fundamental Research, Post Bag 3, Ganeshkhind, Pune
411007, India\\
$^{3}$Inter University Centre for Astronomy and Astrophysics,  Post Bag 4,
Ganeshkhind, Pune 411007, India}

\begin{document}

\maketitle

\label{firstpage}

%%%%%%%%%%%%%%%%%%%%%%%%%%%%%%%%%%%%%%%%%%%%%%%%%%%%%%%%%%%%%%%%%%%%%%

\begin{abstract}

We combine UV/Optical/near-IR/mid-IR data on a sample of $\sim$240
S0 galaxies to examine various star formation related
processes in them.  We split the sample into bright and faint
S0 galaxies based on their $K$ band luminosity.  Comparing the FUV-NUV
versus NUV-K color-color diagram with a Simple Stellar Population
(SSP) model shows that ellipticals and bright S0 galaxies are
dominated by a stellar population of age $> 10^9$ years while faint
S0 galaxies may contain stars as young as $10^8$ years,
providing evidence for relatively recent star formation activity. The
strength of the 4000 \AA\ break is also systematically higher in
brighter S0 galaxies, again indicating the presence of an old stellar
population. Their mid-IR colours indicate that bright S0 
colours are like those of ellipticals while faint S0  colours
are more like spirals. All these observations are consistent with a
scenario in which low-luminosity S0  galaxies likely formed by
the stripping of gas from the discs of late-type spiral galaxies,
which in turn formed their pseudo bulges through secular evolution
processes, possibly involving multiple episodes of star formation. On
the other hand, more luminous S0 galaxies likely formed the bulk of
their stars at early epochs, similar to the star formation in
elliptical galaxies, and are characterised by an old coeval stellar
population and classical bulges.

\end{abstract}

%%%%%%%%%%%%%%%%%%%%%%%%%%%%%%%%%%%%%%%%%%%%%%%%%%%%%%%%%%%%%%%%%%%%%%

\begin{keywords}

galaxies: elliptical and lenticular -   fundamental parameters
galaxies: photometry - structure - bulges 
galaxies: formation - evolution
galaxies: star formation
\end{keywords}

%%%%%%%%%%%%%%%%%%%%%%%%%%%%%%%%%%%%%%%%%%%%%%%%%%%%%%%%%%%%%%%%%%%%%%

\section{Introduction}

Lenticular (S0) galaxies are characterised by the presence of a
central bulge and disk and the absence of spiral arms. Multiwavelength
observations over the decades have not led to any broad consensus
about their origins, properties, and evolution, although they are
often thought to have formed most of their stars early in the history
of the universe and to have evolved relatively passively since
then. Cosmological simulations of galaxy formation appear to show
  a "two-phase" character for galaxy formation with a rapid early
  phase at $z \gtrsim 2$ during which ``in situ'' stars are formed
  within the galaxy from infalling cold gas followed by an extended
  phase since $z \lesssim 3$ during which ``ex situ'' stars are
  primarily accreted \nocite{oser10}({Oser} {et~al.} 2010). In the widely accepted paradigm
of hierarchical galaxy formation, major mergers play an important role
in the galaxy formation process for early type galaxies. In this
picture, current theory predicts that stellar evolution returns
substantial quantities of gas to the interstellar medium; most is
ejected from the galaxy, but significant amounts of cool gas might be
retained \nocite{abadi99,kormendy12,elichemoral12}({Abadi}, {Moore} \& {Bower} 1999; {Kormendy} \& {Bender} 2012; {Eliche-Moral} {et~al.} 2012). This simple picture
is complicated by the need to properly account for the distribution of
merger orbital parameters, gas fractions, and the full efficiency of
merger-induced effects as a function of mass ratio of the merging
galaxies \nocite{hopkins10}({Hopkins} {et~al.} 2010).

The remaining gas may form new stars at different epochs over the
lifetime of the galaxy. Such star formation is however, too weak to
affect the optical light of the galaxy, which continues to be
dominated by the old stellar population and appears "red and dead'' in
optical imaging and spectroscopy (e.g.,
\nocite{bothun90,schweizer92,munn92}{Bothun} \& {Gregg} (1990); {Schweizer} \& {Seitzer} (1992); {Munn} (1992)). 

S0 galaxies occupy an important position in Hubble's tuning fork
diagram \nocite{hubble36}({Hubble} 1936), where they are placed in between ellipticals
and spirals implying that S0 galaxies have properties that are
intermediate between these two classes. Although the view that they
are relatively gas-poor still survives (e.g., \nocite{tomita00}{Tomita} {et~al.} (2000)), it
has become clear that in general ∼75\% - 80\% of them have some dust
and gas (e.g.,
\nocite{knapp85,wardle86,vandokkum95,ferrari99,welch03}{Knapp}, {Turner} \& {Cunniffe} (1985); {Wardle} \& {Knapp} (1986); {van Dokkum} \& {Franx} (1995); {Ferrari} {et~al.} (1999); {Welch} \& {Sage} (2003)). These
differences in neutral and molecular gas content and gas supply are an
essential factor to explain the differences in galaxy formation and
evolution scenarios between different morphologies of galaxies. HI,
FIR and CO observation of S0 galaxies have demonstrated that many of
these objects have an active and cold interstellar medium though
usually in much smaller quantities compared to
spirals. \nocite{balick76}{Balick}, {Faber} \& {Gallagher} (1976) first reported the discovery of atomic
hydrogen in S0 galaxies. \nocite{vandriel91}{van Driel} \& {van Woerden} (1991) observed distributions and
kinematics of HI gas in gas-rich S0 galaxies to study various
scenarios for the origin and evolution of gas in such galaxies. CO
emission from molecular gas was detected in 78\% of field S0 galaxies
in the survey by \nocite{welch03}{Welch} \& {Sage} (2003).

Nearly 60\% of nearby S0 galaxies were detected by IRAS in its 60 $\mu$m
and 100 $\mu$m \nocite{knapp89}({Knapp} {et~al.} 1989) bands indicating the presence of dust. Gas,
dust and small amount of star formation were detected by \nocite{temi09}{Temi}, {Brighenti} \& {Mathews} (2009)
in their sample of early-type galaxies observed with the Spitzer
telescope. The Atlas3D SAURON spectroscopy of early-type galaxies
highlights these results in detail and has detected optical emission
lines and recent star formation in 75\% of its sample galaxies which
included many S0 galaxies \nocite{sarzi06,davis11,young11}({Sarzi} {et~al.} 2006; {Davis} {et~al.} 2011; {Young} {et~al.} 2011). 

These multiwavelength studies of S0 galaxies over the last three decades or so,
indicate that there is great variation in the gross observable
properties of these objects. An important question then is whether these
observed variations in gas and dust content and star formation history
are consistent with {\it all} S0  galaxies having had a similar
formation history.

\nocite{barway07,barway09}{Barway} {et~al.} (2007, 2009) have presented evidence to support the view
that the formation history of S0 galaxies follows two very different
routes. Which route is taken seems to depend primarily upon the {\it
  luminosity} of the galaxy, although the environment also plays a
role. According to this view, luminous S0 galaxies are likely to have
formed their stars at early epochs through major mergers or rapid
collapse followed by rapid star formation, similar to the formation of
elliptical galaxies \nocite{aguerri05}({Aguerri} {et~al.} 2005). The bulges of such S0 galaxies
are more likely {\it classical bulges} \nocite{kormendy04}({Kormendy} \& {Kennicutt} 2004) sharing most
of their stellar and dynamical properties with low and medium mass
elliptical galaxies; the most massive ellipticals seem to be different
\nocite{gadotti09}({Gadotti} 2009). These luminous S0 galaxies with classical bulges
were studied by Hubble, and placed in close proximity to ellipticals
on his tuning fork diagram. On the other hand, low-luminosity S0
galaxies likely formed by the quenching of star formation due to
stripping of gas from the bulges and discs of late-type spiral
galaxies through galaxy interactions or by motion in a dense
environment \nocite{salamanca06, bedregal06, barr07}({Arag{\'o}n-Salamanca},  {Bedregal} \& {Merrifield} 2006; {Bedregal}, {Arag{\'o}n-Salamanca}, \&  {Merrifield} 2006; {Barr} {et~al.} 2007). This fading
scenario may not be applicable for the most massive (and therefore
luminous) S0 galaxies harbouring populous globular clusters
\nocite{sanchezjanssen12}({S{\'a}nchez-Janssen} \& {Aguerri} 2012).

The progenitor spiral galaxies of faint S0s, in turn, likely formed
their {\it pseudo bulges} through secular evolution processes induced
by bars. The bar formation could be episodic; bars induce gas flows
and vigourous star formation \nocite{ellison11,coelho11}({Ellison} {et~al.} 2011; {Coelho} \& {Gadotti} 2011). The bars may
subsequently be disrupted due to the combined effects of central mass
concentration and gravity torques with the mass of the central
concentration being an important parameter
\nocite{bournaud05,athanassoula05}({Bournaud}, {Combes} \&  {Semelin} 2005; {Athanassoula}, {Lambert}, \&  {Dehnen} 2005). The disruption of the bar would
tend to stop the gas inflow to the central regions thereby quenching
star formation. The process may then repeat over galaxy dynamical
timescales \nocite{carollo99,gadotti11,scannapieco12}({Carollo} 1999; {Gadotti} 2011; {Scannapieco} \& {Athanassoula} 2012). Recent work
  has shown that this simple picture may not be correct; the role of
  gas feedback on the weaking of the stellar bar is likely to be
  complicated \nocite{berentzen07}({Berentzen} {et~al.} 2007).  The simulations presented in
  \nocite{athanassoula13}{Athanassoula}, {Machado} \&  {Rodionov} (2013) do not see the destruction or dissolution of
  a bar, which in turn strongly argues for long-lived bars, at least
  in isolated galaxies. Long lived bars are also found in the
  cosmological simulations presented in \nocite{kraljic12}{Kraljic}, {Bournaud} \&  {Martig} (2012).

If this picture is broadly correct (given the uncertainties in bar
longevity), there must be signatures of the formation mechanism
imprinted in the various observable parameters of the galaxy.  Indeed,
such imprints have been seen in the light profile (as correlated bulge
disc-sizes) by \nocite{barway07,barway09}{Barway} {et~al.} (2007, 2009) and the presence of kinematic
structures, such as a stellar bar
\nocite{barway11,vandenbergh12,skibba12}({Barway}, {Wadadekar} \&  {Kembhavi} 2011; {van den Bergh} 2012; {Skibba} {et~al.} 2012). There should also be imprints
on stellar kinematics (as traced by 3D spectroscopy) and in stellar
populations (as traced by broadband colours). The expectation is that
luminous S0 galaxies should have colours characteristic of old stellar
bulges (like those found in most elliptical galaxies), while faint S0
galaxies should have colours indicating multiple episodes of star
formation which would result in a mixed stellar population with a
variety of ages, like those found in spiral galaxies.

In this paper, we combine near and far UV data from the Galaxy
Evolution Explorer (GALEX), optical data from the Sloan Digital Sky
Survey (SDSS), near-IR data from the 2Micron All Sky Survey (2MASS),
and mid-infrared data from the Wide-Field Infrared Survey Explorer
(WISE; \nocite{wright10}{Wright} {et~al.} (2010)) to investigate differences in stellar
populations of a sample of nearby S0 galaxies as a function of
luminosity. We present evidence for a significantly enhanced
probability of recent star formation in faint S0 
galaxies. Throughout this paper, we use the standard concordance
cosmology with $\Omega_M= 0.3$, $\Omega_\Lambda= 0.7$ and $h_{100}=
0.7.$

%%%%%%%%%%%%%%%%%%%%%%%%%%%%%%%%%%%%%%%%%%%%%%%%%%%%%%%%%%%%%%%%%%%%%%
\begin{figure}
\centering
\includegraphics[scale=0.5]{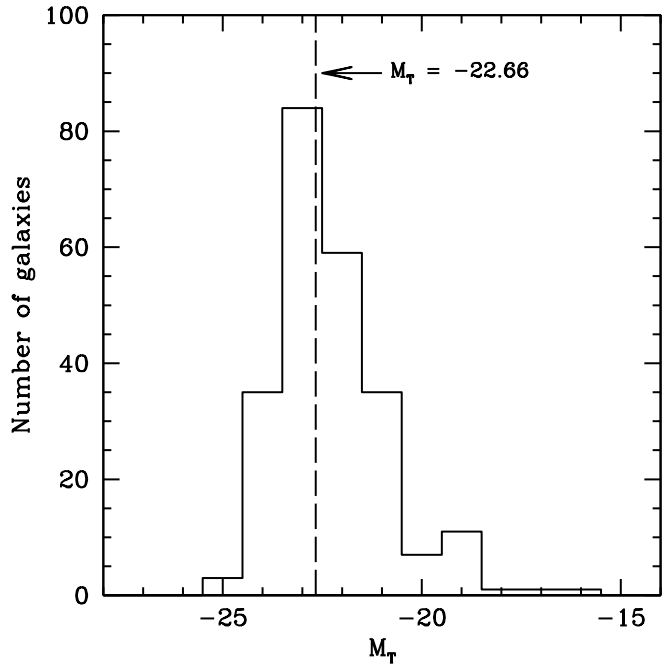}
\caption{Distribution of total absolute AB magnitude (M$_T$) of our
S0  galaxy sample in the $K$ band. The vertical dashed line
corresponds to total absolute magnitude M$_T$ = -22.66, which we use
to divide low- and high-luminosity S0 galaxies.}
\label{f1}
\end{figure}
%%%%%%%%%%%%%%%%%%%%%%%%%%%%%%%%%%%%%%%%%%%%%%%%%%%%%%%%%%%%%%%%%%%%%%

\section{The sample and data}

For this work, we made use of the sample of S0 galaxies constructed by
\nocite{barway11}{Barway} {et~al.} (2011) which has 371 galaxies obtained by cross-matching the
Uppsala General Catalog of Galaxies (UGC; \nocite{nilson73}{Nilson} (1973)) with the
Sloan Digital Sky Survey (SDSS) Data Release 7 (DR7;
\nocite{abazajian09}{Abazajian} {et~al.} (2009)) and 2MASS \nocite{skrutskie06}({Skrutskie} {et~al.} 2006). These galaxies all
have a total apparent blue magnitude brighter than $m_B= 14$. This
sample, while not complete, is a fair representation of the S0 galaxy
population in the local Universe and has a statistically meaningful
number of galaxies spanning a large range of luminosities from the
field as well as group/cluster environments \nocite{barway11}({Barway} {et~al.} 2011). For all
371 galaxies, we have data from SDSS in five bands ($u, g, r, i, z$)
and from 2MASS in the $J$, $H$ and $K$ bands.

For the present work, we cross-matched these 371 galaxies with the UV
imaging data available in the GALEX GR6 data release.  We found a
total of 243 galaxies for which data exist in the FUV and NUV bands of
GALEX. One galaxy, UGC 9094, had a very noisy image in both FUV and
NUV GALEX bands and has been excluded from further analysis. One
should note that the data in the GALEX archive are sourced from a
number of GALEX surveys which includes AIS (all sky, shallow), MIS
(medium), DIS (deep), and NGS (nearby galaxies). The sensitivity
limits are therefore not uniform. However, for the big, bright, nearby
galaxies of our sample, the GALEX imaging has reasonable signal to
noise even in the shallow survey. 

To verify the morphological classification of S0 galaxies in this
sample as listed in the UGC catalog, we performed a visual
classification of these 242 galaxies with good GALEX data. We used two
methods to identify galaxy morphology 1. We examined the color JPEG
images as provided on the SDSS website and 2. We produced and examined
higher signal-to-noise ratio ($S/N$) images made using the prescription
of \nocite{lisker06}{Lisker}, {Grebel} \& {Binggeli} (2006), by co-adding the SDSS $g$, $r$ and $i$ bands.

The visual inspection of galaxy images was performed by two of the
authors independently. A final classification was made after
discussing any inconsistent cases. We found five galaxies (UGC 1157,
UGC 4596, UGC 4963, UGC 5638, UGC 8204) with easy to detect spiral
arms misclassified as S0s. We exclude these from our
sample.  A few other galaxies with very weak asymmetries in structure,
indicative of spirals structure that has faded have been left in, in
the sample. Also, identifying ellipticals that are misclassified as
S0s in the UGC catalog is a difficult exercise. Only a handful
of objects (5 galaxies) fall into this category, but the consistency
in classification with the two methods is weak. We have therefore
chosen to leave these in, as well. This is not surprising because
S0s with a very weak, face on disk appear very much like
ellipticals even with the higher dynamic range of modern digital
surveys. We note that since the number of misclassified objects is
small relative to the size of the sample, a somewhat different
methodology for identifying contaminants would not affect any result
presented in this paper.

We use this sample of 237 galaxies for the analysis presented in this
paper. The FUV and NUV magnitudes measured by the GALEX pipeline are
quoted in the AB magnitude system of \nocite{oke83}{Oke} \& {Gunn} (1983). We corrected the
apparent UV magnitudes for Galactic extinction using the
\nocite{schlegel98}{Schlegel}, {Finkbeiner} \&  {Davis} (1998) reddening maps and assuming the extinction law of
\nocite{cardelli89}{Cardelli}, {Clayton} \&  {Mathis} (1989) and \nocite{wyder05}{Wyder} {et~al.} (2005). SDSS model magnitudes are
already in the AB system and we correct them for Galactic extinction
in the same manner as the UV magnitudes. In order to have all
magnitudes consistently in the AB system, we also converted the 2MASS
K band Vega magnitudes to the AB system by adding 1.84 (as suggested
by \nocite{munozmateos09}{Mu{\~n}oz-Mateos} {et~al.} (2009) and \nocite{cohen03}{Cohen}, {Wheaton} \& {Megeath} (2003)). The converted K band
magnitudes are also corrected for Galactic extinction. All 237
galaxies have mid-IR flux measurements in the 3.4, 4.6 and 12 micron
bands of the WISE satellite \nocite{wright10}({Wright} {et~al.} 2010) which recently carried out
an all sky survey at four mid-IR bands. Unless mentioned otherwise,
all flux measurements are from calibrated source catalogs from the
four wavebands that we use. No $k$-corrections are required as our
galaxies are at a very low median redshift of $z \sim 0.01$. In
Figure~\ref{f1}, we show the distribution of total absolute magnitude
$M_T$ in the $K$ band for galaxies in the sample. We divide the
sample into less luminous and more luminous groups, using $M_T =
-22.66$ as a boundary for absolute magnitudes in the AB system. For
convenience, we use the words {\it bright} and {\it faint} to
designate the more and less luminous groups respectively. The division
line between the two groups corresponds to $M_T = -24.5$ in the Vega
system that we have used in previous papers. With this luminosity
division, 103 galaxies (43\%) are in the bright category while the
remaining 134 (57\%) are in the faint category. Our results do not
critically depend on small (∼0.5 mag) shifts in the dividing
luminosity. Since there is a uniform shift for all galaxies while
moving from the Vega system to the AB system, there is no change in
the fraction of S0 galaxies in the bright and faint groups after the
shift is applied.

%%%%%%%%%%%%%%%%%%%%%%%%%%%%%%%%%%%%%%%%%%%%%%%%%%%%%%%%%%%%%%%%%%%%%%
\begin{figure}
\centering
\includegraphics[scale=0.5]{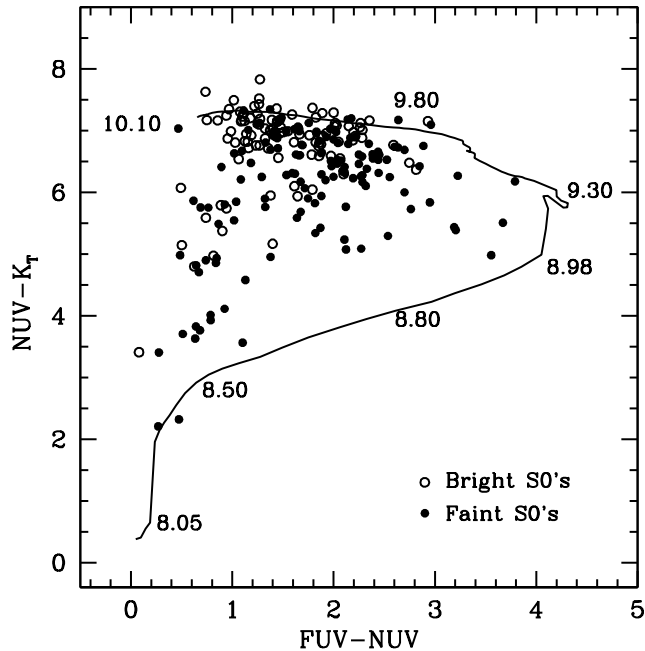}
\caption{Colour-colour diagram for NUV- K$_T$ versus FUV-NUV for bright and
  faint S0 galaxies. The curve is for a Simple Stellar Population
  (SSP) with solar metallicity, Salpeter IMF and Charlot \& Bruzual
  (2007) models. Numbers along the curve indicate the logarithm of the
  age of the SSP. Clearly, bright S0 galaxies are clustered in the
  region of colours space corresponding to an SSP of age $> 10^9$ year
  while fainter S0 galaxies show much greater scatter.}
\label{f2}
\end{figure}
%%%%%%%%%%%%%%%%%%%%%%%%%%%%%%%%%%%%%%%%%%%%%%%%%%%%%%%%%%%%%%%%%%%%%%
\begin{figure}
\centering
\includegraphics[scale=0.5]{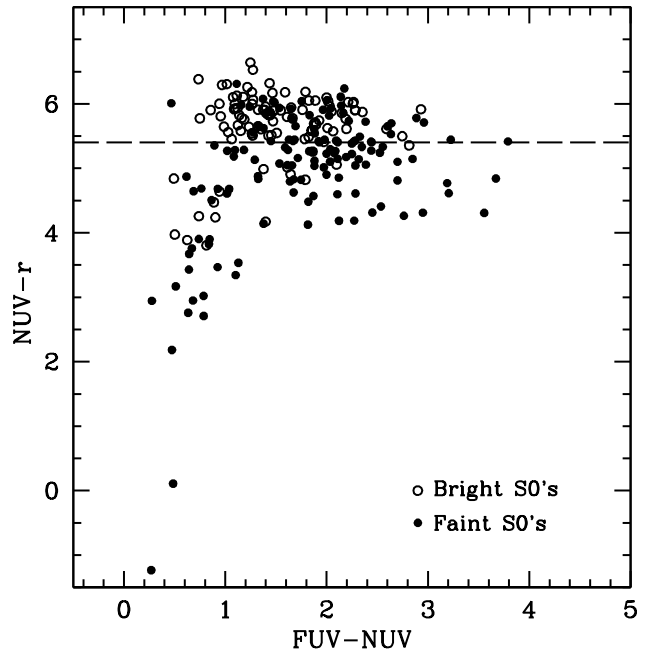}
\caption{Colour-colour diagram for $NUV-r$ versus $FUV-NUV$ for bright and
  faint S0 galaxies. The horizontal dashed line corresponds to the colour
criterion defined by Schawinski et al. 2006 to identify galaxies with {\it
recent star formation}(RSF). Galaxies with RSF lie below the line
in the figure.}
\label{f3}
\end{figure}
%%%%%%%%%%%%%%%%%%%%%%%%%%%%%%%%%%%%%%%%%%%%%%%%%%%%%%%%%%%%%%%%%%%%%%

\begin{figure}
\centering
\includegraphics[scale=0.5]{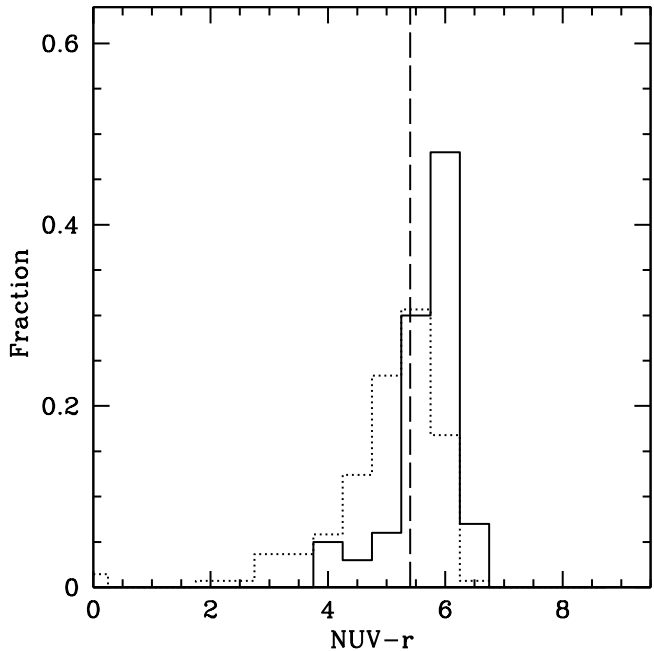}
\caption{\bf Distribution of $NUV-r$ colour for bright (solid line) and
  faint (dotted line) S0 galaxies. The vertical dashed line
  corresponds to the colour criterion defined by
  Schawinski et al. 2006. Galaxies with RSF lie to the left of the
  vertical line in the figure.}
\label{f4}
\end{figure}

\section{Analysis}

\subsection{GALEX-SDSS-2MASS colour-colour diagrams}

Regions of recent star formation in galaxies whose light is dominated
by hot, young stars are well traced by imaging from the Galaxy
Evolution Explorer (GALEX) \nocite{gildepaz07}({Gil de Paz} {et~al.} 2007) telescope. On the other
hand, the near-infrared 2MASS images trace the old stellar population
of galaxies which is dominated by low mass stars. The near-UV (NUV)
detector on GALEX is extremely sensitive to the presence of a young
stellar population. With it, we can detect small mass fractions of
1-3\% of young stars formed within the last billion years
\nocite{schawinski06}({Schawinski} {et~al.} 2006). By plotting a GALEX-2MASS colour-colour diagram
with NUV as the common filter, we can easily separate galaxies with
predominantly young and old stars. The presence of variable amounts of
dust in galaxies can complicate this simple picture somewhat, but the
broad trends are still apparent. In Figure \ref{f2}, we combine GALEX
and 2MASS flux measurements of the galaxies in our sample to plot a
colour-colour diagram (FUV-NUV vs. NUV- K$_T$) with open circles
representing the bright S0 galaxies and dots representing the faint S0
galaxies. The curve shows the age evolution of a Simple Stellar
Population (SSP) with solar metallicity and a Salpeter IMF constructed
using the Charlot \& Bruzual (2007) models. These models are an
improved but unpublished version of the \nocite{bruzual03}{Bruzual} \& {Charlot} (2003)
models. Numbers along the curve indicate the logarithm of the age of
the SSP. Colours on both axes effectively separate old ($>1$ Gyr)
stellar populations from younger ones.  The bulk of the bright S0
galaxies occupy the region of colour-colour space corresponding to an
SSP of age $> 10^9$ years. This indicates that the main stellar
population in such galaxies consists of old, low mass stars. Such a
population is known to predominate in elliptical galaxies. These stars
may all have formed within the same galaxy or dry mergers of multiple
elliptical/S0 progenitors may be involved.  This indicates a close
correspondence between bright S0 galaxies and elliptical
galaxies. Such a result is not at all surprising; many studies of
stellar populations in S0 galaxies over the decades have found a
predominantly old stellar population. Indeed, Hubble's original
placement of S0 galaxies in the tuning fork diagram was motivated by
the close similarities in optical colours between S0 galaxies and
ellipticals.

One bright S0 galaxy lies away from the main group towards the lower
left side of the figure. This is UGC 1597, which is a Seyfert galaxy
with an active core and a disturbed appearance in the SDSS image,
indicative of a recent merger. These peculiarities explain its
extreme blue colours.  The SDSS fiber spectrum of the center of this
galaxy shows very strong $H\alpha$ emission indicating that this is a
strongly star forming galaxy. We exclude this galaxy from the sample
that is analysed in subsequent subsections.

Faint S0 galaxies show a much higher scatter in the figure. More than
half the galaxies lie well away from the region populated by bright
S0 galaxies. Their wide scatter in the colour-colour diagram, and
relative to the SSP curve, indicates that their stellar populations
cannot be explained by a single episode of instantaneous star
formation; their histories must be more complex. We would like to note
here that not all faint S0 galaxies have such complex star formation
histories; a significant fraction lies in the region occupied by
bright S0 galaxies and is reasonably well characterised by a single 
stellar population of advanced age.

UV-Optical colours can also be used to identify galaxies with a young
stellar population. \nocite{schawinski06}{Schawinski} {et~al.} (2006) use a $(NUV-r) < 5.4$ colour
criterion to identify galaxies with {\it recent star formation
}(RSF). In Figure~\ref{f3}, we plot the $NUV-r$ colour versus the
$FUV-NUV$ colour for bright and faint S0 galaxies. Above the line, 83
out of 135 galaxies are bright (61\%); this drops to 17 out of 102
(17\%) in the RSF region below the line. In Figure~\ref{f4}, we
show the distribution of $NUV-r$ colour for the bright and faint
galaxies. Again, there is a clear indication that faint S0 galaxies
are more likely to have {\it recent star formation} than bright ones.

%%%%%%%%%%%%%%%%%%%%%%%%%%%%%%%%%%%%%%%%%%%%%%%%%%%%%%
\begin{figure}
\centering
\includegraphics[scale=0.5]{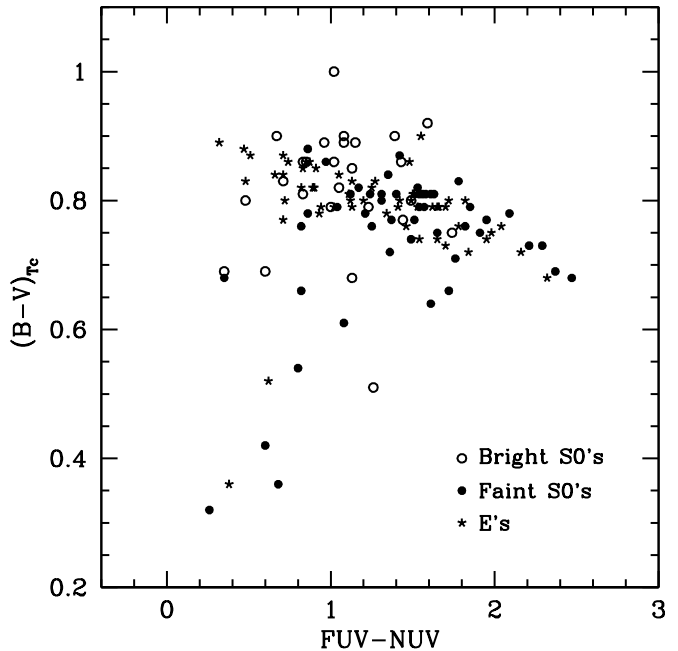}
\caption{Colour-colour diagram for the Donas et al. (2007) galaxy
  sample. Elliptical galaxies are denoted by stars. Bright and
  faint S0 galaxies are denoted by open circles and dots
  respectively.}
\label{f5}
\end{figure}
%%%%%%%%%%%%%%%%%%%%%%%%%%%%%%%%%%%%%%%%%%%%%%%%%%%%%%

Such "star-forming'' S0 galaxies have been noted before in the
literature although the dependence on luminosity was not
discovered. \nocite{donas07}{Donas} {et~al.} (2007) studied a sample of elliptical and
S0  galaxies drawn from the RC3 catalog. In a colour-colour
diagram of $B-V$ (taken from the RC3) versus $FUV-NUV$, they noted a
larger scatter for S0  galaxies; they speculated that this
reflected the presence of low-level star formation. In
Figure~\ref{f5}, we show the colour-colour diagram for the
\nocite{donas07}{Donas} {et~al.} (2007) sample. This figure is similar to Figure. 2 of their
paper. We have, in addition, used 2MASS K-band flux measurements of
all the galaxies in their sample to separate faint and bright
S0 galaxies. The fainter S0 galaxies show a higher scatter than the
brighter ones, which are well mixed with the elliptical
population. The two anomalous elliptical galaxies with blue colours are
starforming ellipticals, which have also been noted as such in
\nocite{donas07}{Donas} {et~al.} (2007).

%%%%%%%%%%%%%%%%%%%%%%%%%%%%%%%%%%%%%%%%%%%%%%%%%%%%%%%%%%%%%%%%%%%%%%

\subsubsection{Environmental dependence}

It would be interesting to see if this difference in star forming
properties between bright and faint S0 galaxies also depends on
whether the galaxy is in the field or in a denser environment. We note
that drawing conclusions based on the current environment of the
galaxy may be misleading because the galaxy may have formed in the
field and then subsequently fallen into a cluster. On the other hand,
a galaxy that appears to be a field galaxy today may have undergone a
merger or interaction in its past that may have altered both its star
formation history and its morphology. Keeping this caveat in mind that
the present environment of a galaxy may be very different from the
environment(s) it has encountered over its lifetime, we use
environmental information for the galaxies in our sample to explore
differences in recent star formation properties as a function of
luminosity and environment. To investigate this issue, we divide the
sample into field and group/cluster environments (for convenience we
use the designation ``cluster'' for both group and cluster
environments) using data from \nocite{tago10}{Tago} {et~al.} (2010), which uses the
friends-of-friends group search method to search for groups in the
SDSS-DR7. Out of 237 S0 galaxies we have used for this study, 64
galaxies are in the field and 173 are members of a cluster, which
reflects the fact that the majority of galaxies located in dense
environments are S0s \nocite{dressler80}({Dressler} 1980) and the majority of S0s are
found in dense environments \nocite{barway11}({Barway} {et~al.} 2011).

\begin{figure}
\centering
\includegraphics[scale=0.5]{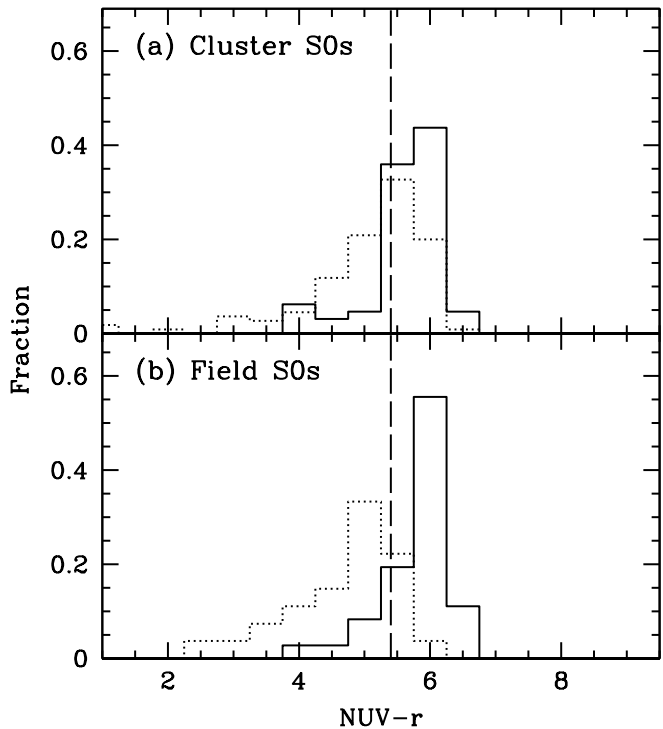}
\caption{\bf Distribution of $NUV-r$ colour for bright (solid line) and
  faint (dotted line)  (a) cluster (b) field S0 galaxies . The
  vertical dashed line corresponds to the colour criterion defined by
  Schawinski et al. 2006. Galaxies with RSF lie to the left of the vertical line in the figure.}
\label{f6}
\end{figure}

In Figure~\ref{f6}, we show the distribution of $NUV-r$ colour for
cluster and field S0 galaxies in our sample. Comparing the two
distributions, it seems that the association between faint and RSF
galaxies and bright and non RSF galaxies is stronger for field S0s
than for S0s in clusters. For galaxies bluer than $NUV-r =4$ in the
field, all except one galaxy are faint. For galaxies without RSF
(those to the right of the line), the bright S0 galaxies outnumber the
faint ones in both field and cluster environments, but the dominance
is much larger for field galaxies.

We note that the number of field galaxies in our sample is
considerably smaller than those in clusters. This implies that the
distribution as plotted is statistically less robust for field
galaxies and the differences with environment may be less significant
than they appear. It seems that the role of environment in influencing
RSF in S0 galaxies is an important secondary effect with the
luminosity as the main differentiator between the two
classes. However, we caution that in attempting to study the
differences between bright and faint S0 galaxies as a function of
environment, we presuppose that the two classes form a homology whose
properties can be meaningfully compared. We have attempted to show
through this work and previous works that bright and faint S0 galaxies
are sufficiently different that a straightforward "apples to apples"
comparison between them is not possible.

%%%%%%%%%%%%%%%%%%%%%%%%%%%%%%%%%%%%%%%%%%%%%%%%%%%%%%%%%%%%%%%%%%
%% \begin{figure}
%% \centering
%% \includegraphics[scale=0.5]{field_env.eps}
%% \caption{Colour-colour diagram for $NUV-r$ versus $FUV-NUV$ for bright and
%%   faint S0 galaxies in the field. The horizontal dashed line is as in Figure~\ref{f3}}
%% \label{field_env}
%% \end{figure}
%% %%%%%%%%%%%%%%%%%%%%%%%%%%%%%%%%%%%%%%%%%%%%%%%%%%%%%%%%%%%%%%%%%%
%% \begin{figure}
%% \centering
%% \includegraphics[scale=0.5]{cluster_env.eps}
%% \caption{Colour-colour diagram for $NUV-r$ versus $FUV-NUV$ for bright and
%%   faint S0 galaxies found in group and cluster environments. The horizontal dashed line is as in Figure~\ref{f3}}
%% \label{cluster_env}
%% \end
%%%%%%%%%%%%%%%%%%%%%%%%%%%%%%%%%%%%%%%%%%%%%%%%%%%%%%%%%%%%%%%%%%%%%%
%%%%%%%%%%%%%%%%%%%%%%%%%%%%%%%%%%%%%%%%%%%%%%%%%%%%%%%%%%%%%%%%%%%%%%

\subsection{$D_{n}(4000)$ index}

\begin{figure}
\centering
\includegraphics[scale=0.5]{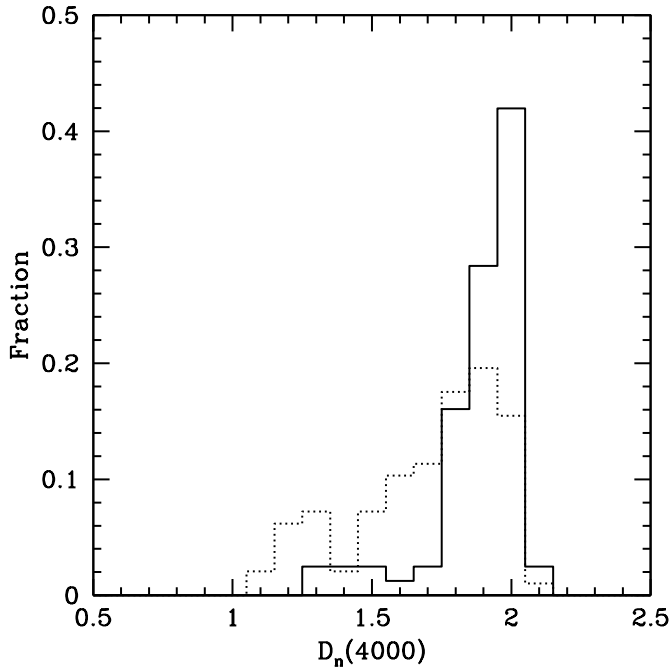}
\caption{Histogram of the $D_{n}(4000)$ index for bright (solid line) and faint (dashed line) S0 galaxies.}
\label{f7}
\end{figure}

Another criterion that can be used to separate old and young stellar
populations in galaxies whose spectra are available, is the strength
of the 4000\AA \ break which arises because of an accumulation of
absorption lines of mainly ionised metals in the atmospheres of old,
low mass stars and by a deficiency of hot, blue stars in galaxies. The
strength of the break is quantified by the $D_{n} (4000)$ index with
higher values of the index seen in old elliptical galaxies without
recent/ongoing star formation. There are several definitions of this
index in the literature; for this work we use a recent definition
provided by \nocite{balogh99}{Balogh} {et~al.} (1999). Of the 371 galaxies in the original
sample with SDSS imaging, 180 galaxies have SDSS spectra available in
the DR9 release \nocite{ahn12}({Ahn} {et~al.} 2012). This rather small number of galaxies
with available spectra is due to the fact that the galaxy sample in
SDSS is limited at the bright end by the fiber magnitude limits, to
avoid saturation and excessive cross-talk in the spectrographs. For
the galaxies with spectra, recent data releases (we used the most
recent DR9) of the SDSS provide measurements of the $D_{n}(4000)$
index in the {\it galspecindx} table. Figure~\ref{f7} shows the
distribution of the $D_{n}(4000)$ index for bright and faint S0
galaxies. Bright S0 galaxies are sharply peaked at around $D_{n}(4000)
= 2$ indicating that most of these objects are similar in their
stellar populations to elliptical galaxies. Faint S0 galaxies, on the
other hand, show a much wider distribution indicating that they have a
mixed population of old and young stars.

%%%%%%%%%%%%%%%%%%%%%%%%%%%%%%%%%%%%%%%%%%%%%%%%%%%%%%%%%%%%%%%%%%%%%%
%%%%%%%%%%%%%%%%%%%%%%%%%%%%%%%%%%%%%%%%%%%%%%%%%%%%%%%%%%%%%%%%%%%%%%

\subsection{WISE mid-IR colour-colour diagram}

\begin{figure}
\centering
\includegraphics[scale=0.41]{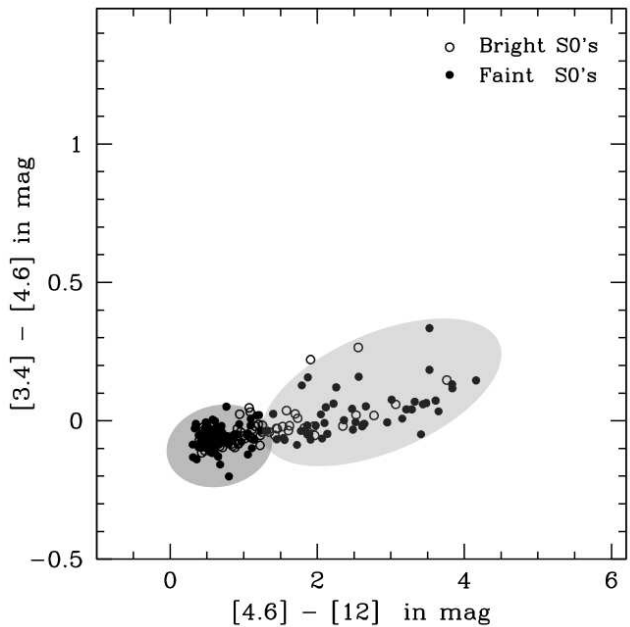}
\caption{Colour-colour diagram for WISE (3.4 - 4.6) micron versus WISE
  (4.6 - 12) micron for bright and faint S0 galaxies. The dark grey
  shaded ellipse is the region of colour colour space occupied by
  elliptical galaxies while the light grey shaded ellipse is the
  region occupied by spiral galaxies.}
\label{wise}
\end{figure}

All sky imaging and source catalogs in 4 mid-infrared bands are now
available from the WISE data archive. Using data from 3 of these
bands, \nocite{wright10}{Wright} {et~al.} (2010) have plotted a color-color diagram (see their
Fig. 12), to indicate the positions of different classes of
astronomical sources in mid-IR colour space. All the galaxies in our
sample have data in the WISE survey; following \nocite{wright10}{Wright} {et~al.} (2010) we
plot a colour-colour diagram of (3.4 - 4.6) micron colour versus (4.6 -
12) micron colour for our sample in Figure~\ref{wise}. The dark grey
shaded ellipse is the region of colour-colour space occupied by
elliptical galaxies while the light grey shaded ellipse is the region
occupied by spiral galaxies. In mid-IR colour space, bright S0
galaxies dominate in the region of elliptical galaxies, while faint S0
galaxies dominate in the region of spiral galaxies. Mid IR colours
seem to strengthen the association between bright S0s and ellipticals
and faint S0s and spirals.

%%%%%%%%%%%%%%%%%%%%%%%%%%%%%%%%%%%%%%%%%%%%%%%%%%%%%%%%%%%%%%%%%%%%%%
%%%%%%%%%%%%%%%%%%%%%%%%%%%%%%%%%%%%%%%%%%%%%%%%%%%%%%%%%%%%%%%%%%%%%%

\section{Conclusion}

In a series of previous papers, \nocite{barway07,barway09,barway11}{Barway} {et~al.} (2007, 2009, 2011) we
have presented evidence to suggest that the two types of S0 galaxies
differentiated by luminosity have fundamentally different formation
scenarios. The two populations show systematic differences in a number
of correlations ($r_e - r_d$, Kormendy relation, photometric plane
etc.) and in their bar fraction. Their bulges retain signatures of
their radically different formation history - those in bright S0
galaxies are overwhelmingly classical bulges; a large fraction of
those in faint S0 galaxies are pseudo bulges.

The analysis of UV-Optical-nearIR-midIR colours of S0 galaxies in the
present paper, when used as simple estimators of their star formation
history, also lends support to the view that bright and faint S0
galaxies are fundamentally different. Stellar populations in bright S0
galaxies resemble those in ellipticals and bulges of early-type
spirals suggesting that they may have formed the bulk of their stars
at early epochs via major mergers or rapid collapse. These galaxies
may also have a smaller amount of recent star formation due to minor
mergers \nocite{kaviraj07}({Kaviraj} {et~al.} 2007), consistent with the hierarchical
paradigm. Just over half the faint S0 galaxies, on the other hand,
seem to have stellar populations that cannot be explained by a simple
stellar population of any age. These galaxies must have had multiple
major episodes of star formation. Such a complex star formation
history is expected when the bulge grows via internal secular
evolution processes with episodic star formation fueled by bar driven
inflows (more likely in the field) or via environment influenced
secular evolution processes such as minor mergers, ram pressure
stripping and galaxy harassment (more common in groups/clusters). In
internal secular evolution, the transformation to a S0 {\it
  appearance} can only happen when the bulge luminosity becomes
comparable to the disk luminosity and the star formation in the spiral
arms becomes sufficiently weak to render them invisible. Both internal
and environment induced processes, when operating on spiral galaxies,
would tend to transform them into S0 galaxies.

$K$ band luminosity is a reasonable proxy for stellar mass which is
dominated by contributions from low mass stars whose emission peaks in
the near-infrared band. It may well be that stellar mass and not
K-band luminosity is a more relevant physical parameter that separates
the two classes of S0 galaxies. Nevertheless, we have chosen to use
$K$ band luminosity in our analysis, since it is a directly measured
quantity from the imaging, unlike the mass whose computation requires
some model dependent assumptions.

It would be interesting to see if kinematic signatures as traced by 3D
spectroscopy would be different for the two galaxy types. We hope to
explore this aspect in a forthcoming paper.

%%%%%%%%%%%%%%%%%%%%%%%%%%%%%%%%%%%%%%%%%%%%%%%%%%%%%%%%%%%%%%%%%%%%%%

\section*{acknowledgements}

We thank the anonymous referee for insightful comments that have
improved both the content and presentation of this paper. YW thanks
Sushruti Santhanam for continuous support, especially during his visit
to the SAAO.

This paper is based upon work supported financially by the National
Research Foundation (NRF). Any opinions, findings and conclusions or
recommendations expressed in this paper are those of the authors and
therefore the NRF does not accept any liability in regard
thereto. This work was supported by a bilateral grant under the
Indo-South Africa Science and Technology Cooperation (UID-76354)
funded by Departments of Science and Technology (DST) of the Indian
and South African Governments.

GALEX (Galaxy Evolution Explorer) is a NASA Small Explorer, launched
in 2003 April. We gratefully acknowledge NASA's support for
construction, operation, and science analysis for the GALEX mission,
developed in cooperation with the Centre National d'Études Spatiales
of France and the Korean Ministry of Science and Technology. 

This research has made use of the NASA/IPAC Extragalactic Database
(NED), which is operated by the Jet Propulsion Laboratory, California
Institute of Technology (Caltech) under contract with NASA. 

Funding for SDSS-III has been provided by the Alfred P. Sloan
Foundation, the Participating Institutions, the National Science
Foundation, and the U.S. Department of Energy Office of Science. 
The SDSS-III web site is http://www.sdss3.org/.

SDSS-III is managed by the Astrophysical Research Consortium for the
Participating Institutions of the SDSS-III Collaboration including the
University of Arizona, the Brazilian Participation Group, Brookhaven
National Laboratory, University of Cambridge, Carnegie Mellon
University, University of Florida, the French Participation Group, the
German Participation Group, Harvard University, the Instituto de
Astrofisica de Canarias, the Michigan State/Notre Dame/JINA
Participation Group, Johns Hopkins University, Lawrence Berkeley
National Laboratory, Max Planck Institute for Astrophysics, Max Planck
Institute for Extraterrestrial Physics, New Mexico State University,
New York University, Ohio State University, Pennsylvania State
University, University of Portsmouth, Princeton University, the
Spanish Participation Group, University of Tokyo, University of Utah,
Vanderbilt University, University of Virginia, University of
Washington, and Yale University.

This publication makes use of data products from the Two Micron All
Sky Survey, which is a joint project of the University of
Massachusetts and the Infrared Processing and Analysis
Center/California Institute of Technology, funded by the National
Aeronautics and Space Administration and the National Science
Foundation. This publication makes use of data products from the
Wide-field Infrared Survey Explorer (WISE), which is a joint project
of the University of California, Los Angeles, and the Jet Propulsion
Laboratory/California Institute of Technology, funded by the National
Aeronautics and Space Administration.

%%%%%%%%%%%%%%%%%%%%%%%%%%%%%%%%%%%%%%%%%%%%%%%%%%%%%%%%%%%%%%%%%%%%%%

%% \bibliography

%%%%%%%%%%%%%%%%%%%%%%%%%%%%%%%%%%%%%%%%%%%%%%%%%%%%%%%%%%%%%%%%%%%%%%

\end{document}